\documentclass[sigconf]{acmart}
\AtBeginDocument{%
  }

\setcopyright{acmlicensed}
\copyrightyear{2025}
\acmYear{2025}
\acmDOI{XXXXXXX.XXXXXXX}
\acmConference[Conference acronym 'XX]{Make sure to enter the correctconference title from your rights confirmation email}{June 03--05,2018}{Woodstock, NY}
\acmISBN{978-1-4503-XXXX-X/2018/06}

\usepackage{caption}
\usepackage{subcaption}
\usepackage{multirow} 
\usepackage{tabularx} 
\newcolumntype{Y}{>{\centering\arraybackslash}X}
\begin{document}

\title{SaviorRec: Semantic-Behavior Alignment for Cold-Start Recommendation}

\author{Yining Yao}
\authornote{Both authors contributed equally to this research.}
\email{yaoyining.yyn@taobao.com}
\affiliation{%
  \institution{Alibaba Group}
  \city{Hangzhou}
  \country{China}}

\author{Ziwei Li}
\authornotemark[1]
\email{jinhe.lzw@taobao.com}
\affiliation{%
  \institution{Alibaba Group}
  \city{Hangzhou}
  \country{China}}

\author{Shuwen Xiao}
\email{shuwen.xsw@alibaba-inc.com}
\affiliation{%
  \institution{Alibaba Group}
  \city{Hangzhou}
  \country{China}}

\author{Boya Du}
\email{boya.dby@taobao.com}
\affiliation{%
  \institution{Alibaba Group}
  \city{Hangzhou}
  \country{China}}

\author{Jialin Zhu}
\email{xiafei.zjl@taobao.com}
\affiliation{%
  \institution{Alibaba Group}
  \city{Hangzhou}
  \country{China}}

\author{Junjun Zheng}
\email{fangcheng.zjj@alibaba-inc.com}
\affiliation{%
  \institution{Alibaba Group}
  \city{Hangzhou}
  \country{China}}

\author{Xiangheng Kong}
\email{yongheng.kxh@alibaba-inc.com}
\affiliation{%
  \institution{Alibaba Group}
  \city{Hangzhou}
  \country{China}}

\author{Yuning Jiang}
\email{mengzhu.jyn@alibaba-inc.com}
\authornote{Corresponding author.}
\affiliation{%
  \institution{Alibaba Group}
  \city{Hangzhou}
  \country{China}}

\renewcommand{\shortauthors}{Yao et al.}

\begin{abstract}
In recommendation systems, predicting Click-Through Rate (CTR) is crucial for accurately matching users with items. To improve recommendation performance for cold-start and long-tail items, recent studies focus on leveraging item multimodal features to model users' interests. However, obtaining multimodal representations for items relies on complex pre-trained encoders, which incurs unacceptable computation cost to train jointly with downstream ranking models. Therefore, it is important to maintain alignment between semantic and behavior space in a lightweight way.

To address these challenges, we propose a Semantic-Behavior Alignment for Cold-start Recommendation framework, which mainly focuses on utilizing multimodal representations that align with the user behavior space to predict CTR. 
First, we leverage domain-specific knowledge to train a multimodal encoder to generate behavior-aware semantic representations.
Second, we use residual quantized semantic ID to dynamically bridge the gap between multimodal representations and the ranking model, facilitating the continuous semantic-behavior alignment.
We conduct our offline and online experiments on the Taobao, one
of the world's largest e-commerce platforms, and have achieved an increase of 0.83\% in offline AUC, 13.21\% clicks increase and 13.44\% orders increase in the online A/B test, emphasizing the efficacy of our method.
\end{abstract}




\maketitle

\section{Introduction}
\begin{figure}[htbp]
  \centering
  \begin{subfigure}{0.5\linewidth}
    \centering
    \includegraphics[width=\linewidth]{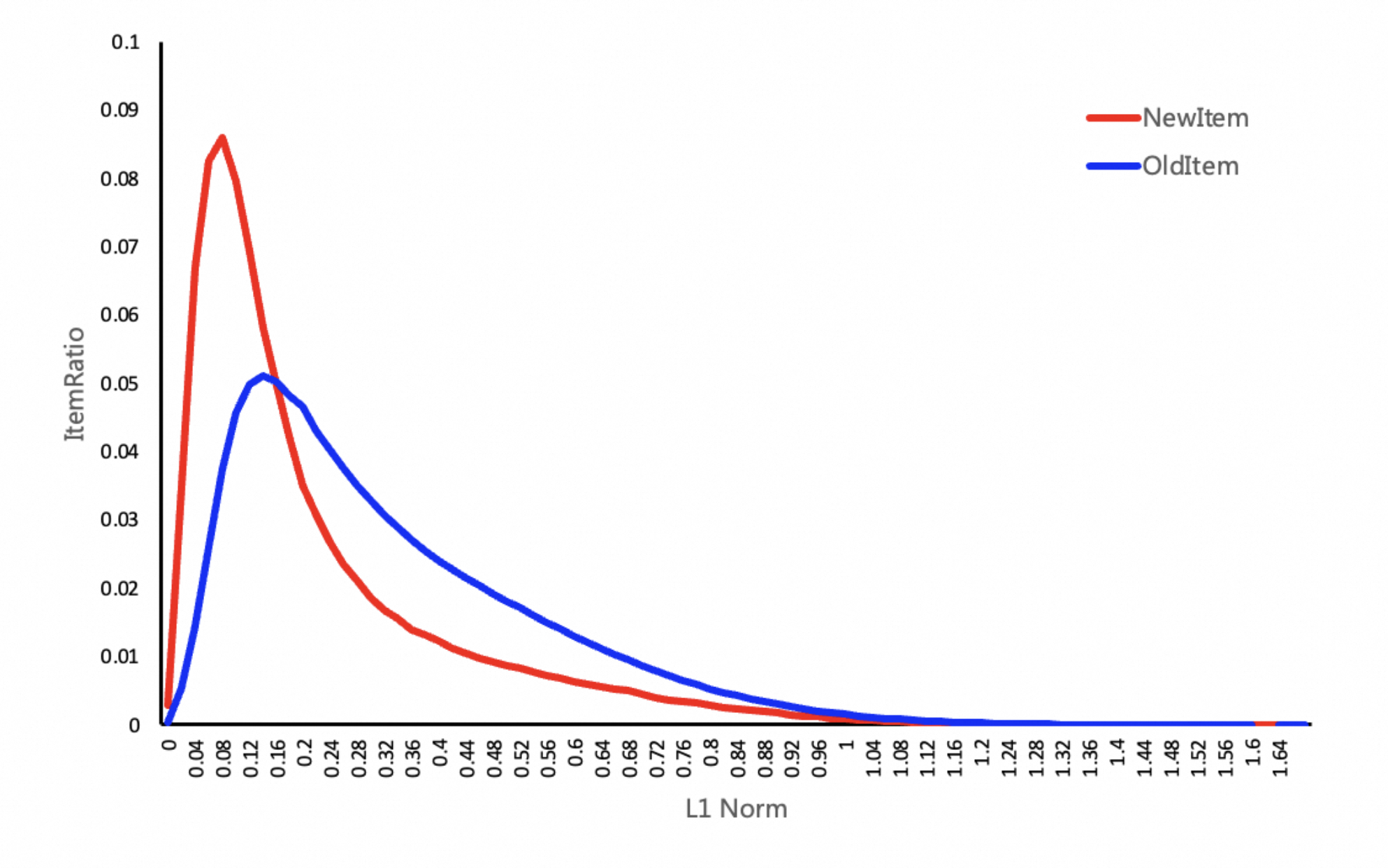}
    \label{fig:itemIDNorm}
  \end{subfigure}
  \hfill
  \begin{subfigure}{0.49\linewidth}
    \centering
    \includegraphics[width=\linewidth]{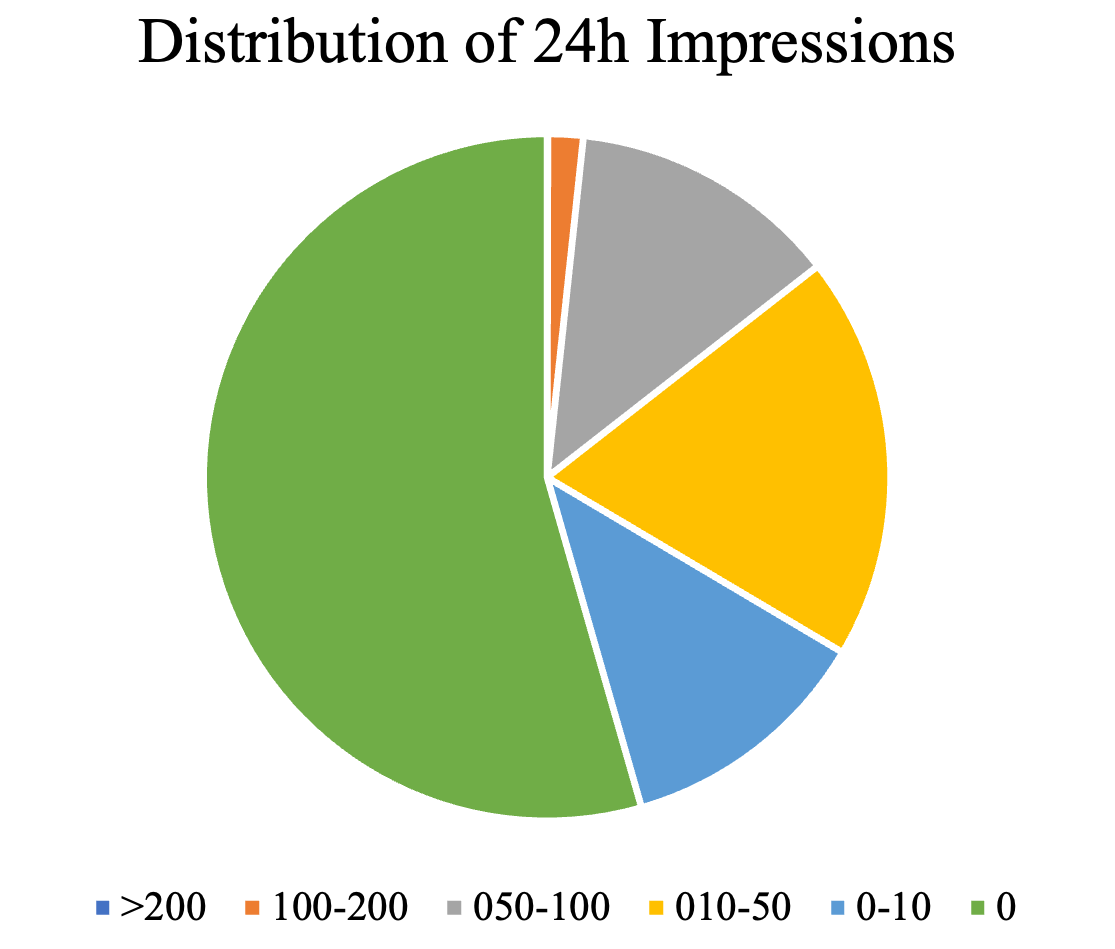}
    \label{fig:24hpv}
  \end{subfigure}
  \caption{Item ID and statistical features distribution of cold-start items in Taobao.}
  \label{fig:feature_distribution}
\end{figure}

In recommendation systems, predicting Click-Through Rate (CTR) is of great importance to accurately match users with items that align with their interests. Current recommendation systems\cite{dcn,din,ytbdnn,eta} primarily rely on features extracted from user-item interaction history to model candidate items and predict CTR, specifically, item ID embeddings, along with statistical features such as historical impression. However, for cold-start and long-tail items, CTR prediction based on these features still faces limitations: ID embeddings are challenging to learn sufficiently because of the lack of training samples, while statistical features are often too sparse, because these items receive few impressions or clicks, as shown in Fig.~\ref{fig:feature_distribution}. Consequently, traditional recommendation systems lack the ability to effectively model less popular items, which potentially leads to constrained sales growth and reduced user experience.

To address the limitation in CTR prediction within cold-start and long-tail scenarios, recent works\cite{m3csr,qarm,mim,bbqrec} mainly focus on integrating multimodal features to represent both candidate item and users interacted item sequence, further enabling semantic modeling of user interests. With the advancement in multimodal representation learning, various pre-trained multimodal foundation models such as VisualBERT\cite{visualbert}, CLIP\cite{clip}, LLaMA\cite{llama}, and ViT\cite{vit} can generate representation embeddings based on modal information including images, text, and audio. These representations can be further utilized by downstream tasks. In recommendation systems, the multimodal features of items are decoupled from user behavior, with little distinction between popular and cold-start items. This allows recommendation systems to overcome the limitations associated with reliance on behavioral information and to achieve better recommendations for cold-start items\cite{fair}.

In recent studies the focus is mostly on two parts: (1) how to extract multimodal features aligning with specific recommendation scenario, and (2) how to integrate multimodal features into recommendation system. For extracting multimodal features, most studies\cite{qarm,pmmrec,deepset} finetune pre-trained multimodal models by contrastive learning. This approach encourages that items close in user behavioral space also have similar multimodal representations, which helps bridge the gap between general pre-trained models and specific recommendation tasks. For integrating multimodal features into recommendation systems, some methods\cite{deepset,mim} capture the semantic user interests from the sequence of items that users have interacted with, then compute the similarity between the interest representation and the multimodal representation of candidate items, further infer whether the user is interested in the candidate item. Other studies employ residual quantization (RQ) method to extract a discretizd, multi-level category-like semantic ID from multimodal feature. \cite{vqrec,tiger,letter,mmgrec,onerec,bbqrec} use generative models widely used in large language models (LLM) to directly predict the semantic ID of the next item that user will likely interact with based on their historical interactions, while \cite{enhancing,unified,spm} map semantic ID to a semantic embedding to replace randomly hashed item ID, reducing the instability caused by irrelevant items sharing the same item embedding.

Although existing methods have made significant progress in leveraging multimodal features to assist cold-start recommendation, there still remain several challenges and limitations.

\textbf{Disjoint between trainable ranking model and fixed multimodal embedding.}

In industrial-scale recommendation systems, where real-time user requests demand timely updates of parameters and feature embeddings, existing methods\cite{deepset,mim} face a critical limitation: they follow a 2-stage schema that first extracts item multimodal embeddings through a large pre-trained encoder, and then feeds these fixed embeddings to downstream ranking models. The complexity of multimodal encoder prevents it from updating in real time, resulting in an increasing gap between multimodal features and dynamically evolving user behavior patterns.

\textbf{Underutilization of multimodal information.}

While some efforts\cite{spm,enhancing} aim to enable trainable multimodal representations with residual quantization (RQ) semantic ID, these methods neglect raw multimodal embeddings, which leads to information loss. Some studies\cite{deepset,chime} measure modal similarity between a candidate item and the user's interaction sequence. However, the lack of interaction between multimodal feature and other behavioral signals limits the full potential of multimodal representations for user interest modeling.


To tackle these challenges, we propose \textbf{S}emantic-Beh\textbf{avior} Alignment for Cold-start \textbf{Rec}ommendation (\textbf{SaviorRec}) framework, to integrate multimodal information into recommendation model. Specifically, our SaviorRec consists of 3 parts. Firstly, \textbf{SaviorEnc} follows a 2-stage paradigm, deriving behavior-aware semantic representations. Secondly, we design a trainable \textbf{Modal-Behavior Alignment (MBA)} block that ensures consistency between the semantic space and the behavior space, while also preserving the important raw multimodal feature. Thirdly, to integrate multimodal features into the ranking model, we propose a \textbf{bi-directional target attention mechanism} between behavior feature and multimodal feature. This design emphasizes the mutual influence between users' interaction and semantic information, and is beneficial for accurate user interest modeling. We conduct our offline and online experiments on Taobao, one of the world's largest e-commerce platforms, and experiment results show that our SaviorRec effectively incorporates the multimodal information into the ranking model and enhances performance in CTR prediction.

In summary, our main contributions are as follows:\begin{itemize}
\item We train a behavior-aware multimodal encoder to obtain multimodal embeddings and semantic IDs.
\item We propose a plug-in module to achieve continuous alignment between semantic and behavior space during the update of ranking model, and further design a bi-directional target attention mechanism to enhance the interaction between behavior signal and semantic information.
\item We conduct offline experiments and ablation study to demonstrate the effectiveness of SaviorRec and further validate its performance through online A/B test, where it achieves significant gains.
\end{itemize}

\begin{figure*}[htbp]
  \centering
  \includegraphics[width=0.99\textwidth]{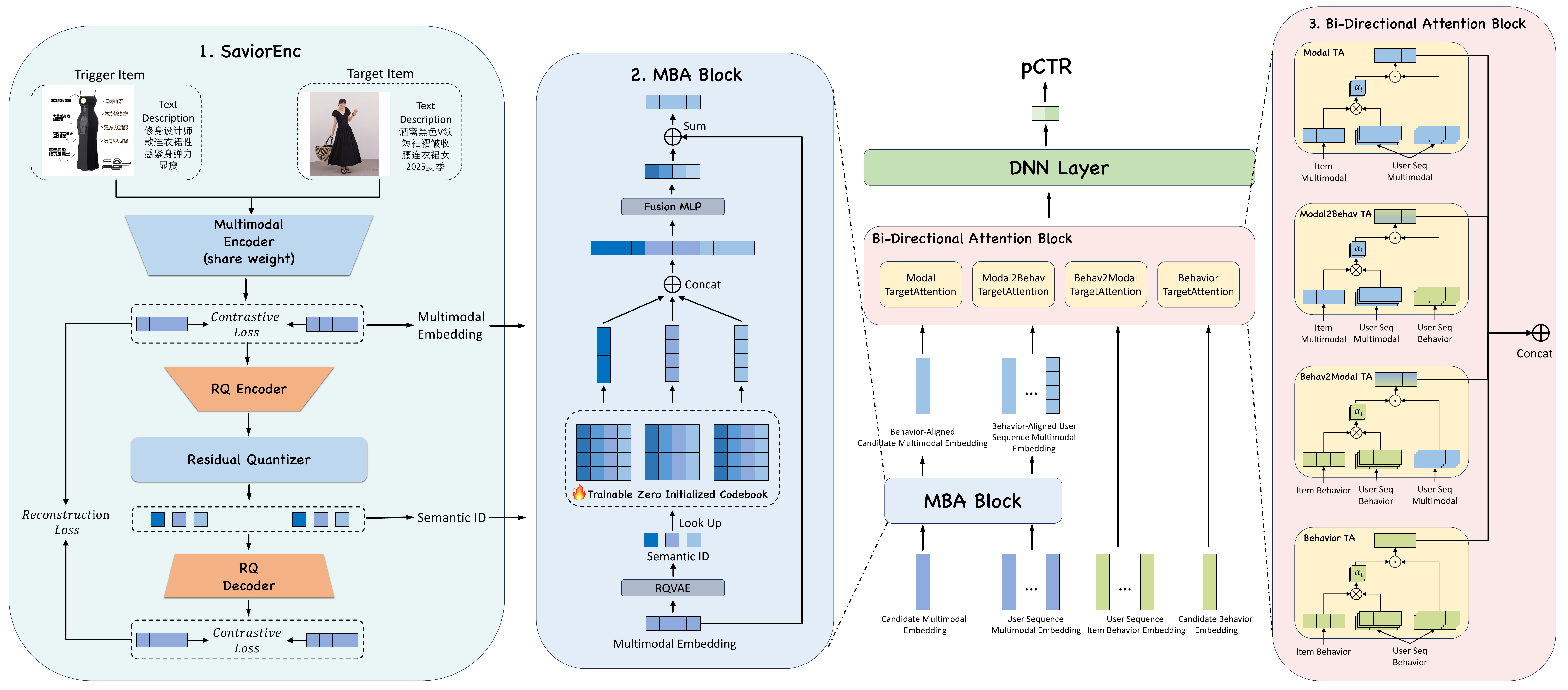}
  \caption{The overview of our SaviorRec framework. SaviorRec mainly consists of three parts. (1) SaviorEnc, (2) MBA block, (3) Bi-directional attention block.}
  \label{fig:model_overview}
\end{figure*}

\section{Related Works}
\subsection{Multimodal Representation Learning}
Recent advances in large-scale pre-training have significantly advanced the ability of models to understand multimodal information. These pre-training paradigms can be broadly categorized into two families: reconstructive/generative and contrastive. 
The generative paradigm trains models to reconstruct masked or missing portions of the input data, thus learning rich, contextualized features. This includes objectives like Masked Language/Image Modeling (MLM/MIM)\cite{bao2022beitbertpretrainingimage,wang2022imageforeignlanguagebeit,kim2021viltvisionandlanguagetransformerconvolution}, where models predict hidden text tokens or image patches based on the surrounding multimodal context. Other approaches frame the task as image captioning or text-to-image generation \cite{yu2022cocacontrastivecaptionersimagetext,wang2022simvlmsimplevisuallanguage,li2022blipbootstrappinglanguageimagepretraining}, forcing the model to understand the deep semantic connections between modalities. While these methods excel at learning an item's objective attributes, their application in recommendation faces a challenge: a model's ability to reconstruct content does not inherently capture user preference.
The contrastive paradigm learns a shared embedding space by aligning positive pairs and pushing apart negative ones. 
This approach \cite{clip,yang2023chineseclipcontrastivevisionlanguage,li2022blipbootstrappinglanguageimagepretraining,jia2021scalingvisualvisionlanguagerepresentation,zhai2023sigmoidlosslanguageimage,sun2023evaclipimprovedtrainingtechniques} has proven highly effective in learning powerful representations from large-scale image-text data. However, a critical gap remains: the general-world alignment learned from Web data often fails to capture the subjective interests of users in e-commerce domain. To address this, our work explicitly uses user co-interaction patterns as the supervisory signal for contrastive tuning, directly embedding behavioral relationships into the semantic space.

\subsection{Multimodal Recommendation}
Introducing multimodal information into recommendation systems enables comprehensive modeling of items, especially in cold-start and long-tail scenarios. Traditional methods\cite{image} simply use image pixel-level features to enhance representation power. However, recommendation models often lack the ability to fully capture these information. Recent studies are based on a two-stage paradigm: a pre-trained large model is finetuned to generate a multimodal embedding vector for each item, which is then consumed by the downstream recommendation model. SimTier\cite{deepset} computes a histogram of the cosine similarity between the candidate item and the user behavior sequence as a multimodal feature. QARM\cite{qarm} uses the quantitative code mechanism to compress multimodal embeddings as semantic IDs for end-to-end training. Though effective, these methods ignore important information contained in original multimodal embeddings. \cite{enhancing,unified,spm} use multimodal semantic ID to replace item ID, avoiding random collision when hashing item to ID embedding. Other methods\cite{onerec,tiger,bbqrec} quantize multimodal features into discrete IDs, and then use a generative encoder-decoder structure to predict the next item.

\section{Methodology}
\subsection{Task Definition and Overview}
\subsubsection{Task Definition for Cold-Start CTR Prediction}
\label{sec:task}

The Click-Through Rate (CTR) prediction task is to predict the probability of a user clicking a candidate item using a ranking model. Specifically, in the cold-start recommendation scenario of Taobao, for each user 
\begin{math}
u \in \mathcal{U}
\end{math} 
, we use a unique user ID
\begin{math}
ID_u 
\end{math}
, user profile 
\begin{math}
P_u 
\end{math}
and items that the user interacted with
\begin{math}
Seq_u = [item_u^1,\ ...,\ item_u^n]
\end{math} to model the user's interest.
And for each candidate item
\begin{math}
i \in \mathcal{I}
\end{math} we construct a series of features including item ID
\begin{math}
ID_i
\end{math},
statistical features extracted from user interactions
\begin{math}
S_i
\end{math}
and multimodal feature
\begin{math}
MM_i
\end{math}.
In addition, we leverage a base CTR score 
\begin{math}
pCTR_{u,i}^{base}
\end{math}
predicted by the model of main recommendation scenario, which, although not very accurate for cold-start and long-tail items, still serves as a valuable reference.

Based on the input features above, the cold-start CTR prediction task can be formulated as:
\begin{equation}
    pCTR_{u,i}=f([ID_u,\ P_u,\ Seq_u],\ [ID_i,\ S_i,\ MM_i],\ pCTR_{u,i}^{base})
\end{equation}
\subsubsection{Method Overview}
The overall structure of our method is shown in Fig.\ref{fig:model_overview}, which takes features described in \ref{sec:task} as inputs and predicts CTR score through a deep neural network. Our method consists of three important parts. SaviorEnc extracts multimodal embedding and residual quantized semantic ID from image and text description of items. Modal-behavior alignment module (MBA) utilizes semantic ID to adjust the multimodal embedding, ensuring the alignment between semantic and behavior space. Bi-directional target attention block extracts user interests from the behavior sequence, and achieves information fusion between behavioral and multimodal features.

\subsection{SaviorEnc}
\label{sec:3_2}

A key challenge in multimodal recommendation is the semantic gap between the pre-training objectives and the user interests in downstream tasks. To address this, we design SaviorEnc, a multimodal encoder following a 2-stage schema that learns multimodal representations directly from user-interaction signals. Stage 1 trains a powerful encoder to extract multimodal embedding and stage 2 employsa RQ-VAE to generate corresponding semantic ID.

To obtain powerful multimodal encoders, we initialize our model with the open-source CN-CLIP \cite{yang2023chineseclipcontrastivevisionlanguage} and perform a domain adaptation fine-tuning, training CN-CLIP on our proprietary product dataset. This serves as the foundation for the behavior-aware encoder.

At stage 1, we construct a set of positive item pairs $\mathcal{P}^+$ by mining frequent co-click patterns $(i,j)$ from user logs. This data-driven approach allows us to capture both intrinsic item similarity and complementarity. Further, we train a contrastive model with the co-click pairs. For each item $i$, we obtain unimodal representations from a vision encoder $f_v(v_i)$ and a text encoder $f_t(t_i)$, which are then fed into a multimodal transformer to capture cross-modal interactions, yielding a fused representation. The output is subsequently mapped to a latent space via an MLP projection head, $g_{proj}(\cdot)$, for the contrastive loss calculation:
\begin{equation}
\mathbf{z}_i = g_{proj}(g([f_v(v_i) ; f_t(t_i)])) \in \mathbb{R}^{d}
\end{equation}
The model is optimized using an InfoNCE loss \cite{oord2019representationlearningcontrastivepredictive}, which maximizes the agreement between positive pairs $(i, j)$ against in-batch negatives. The loss for the item $i$ can be formulated as:
\begin{equation}
\mathcal{L}_{i} = -\log \frac{\exp(\text{sim}(\mathbf{z}_i, \mathbf{z}_j)/\tau)}{\exp(\text{sim}(\mathbf{z}_i, \mathbf{z}_j)/\tau) + \sum_{k \neq i,j} \exp(\text{sim}(\mathbf{z}_i, \mathbf{z}_k)/\tau)}
\end{equation}
where $\text{sim}(\cdot, \cdot)$ is cosine similarity and $\tau$ is temperature. Through these training objectives, we obtain a dense multimodal feature $\mathbf{z}_i$ for each item.  The resulting embedding space is semantically aligned with user behaviors, where co-interacted items are mapped closer together.

While the multimodal representation $\mathbf{z}_i$ is effective, it still presents a challenge for ranking models, as these large, frozen embeddings cannot be fine-tuned. To enable dynamic updating and continuous behavior alignment of the multimodal embeddings detailed in \ref{sec:3_3}, at stage 2 we discretize $\mathbf{z}_i$ into a sequence of semantic IDs, inspired by \cite{rqvae,rqaudio}.

Specifically, we employ a residual quantized variational autoencoder (RQ-VAE) \cite{zeghidour2021soundstreamendtoendneuralaudio} which learns to map a continuous vector to a sequence of discrete codes $\mathbf{c} = (c_i^1, c_i^2, ..., c_i^L)$ from $L$ different codebooks $C = \{C_1, ..., C_L\}$. The quantization is performed residually. The first quantizer finds the closest code for the input vector, and the second quantizer finds the code for the residual result of the first. This process continues iteratively for subsequent quantizers.

\begin{align}
&\mathbf{r}_i^1 = \text{Enc}(\mathbf{z}_i) \notag \\
& c_i^l = \text{argmin}_c \|\mathbf{r}_i^l - \mathbf{C}_l^c\|_2 \quad \text{for } l=1, ..., L\\
&\mathbf{r}_i^{l+1} = \mathbf{r}_i^l - \mathbf{C}_l^{c^l_i} \notag
\end{align}

The sequence of indices of the vectors $\{\mathbf{C}_l^{c^l_i}\}_{l=1}^L$ forms the semantic ID for item $i$. The reconstructed vector is $\hat{\mathbf{z}}_i = \text{Dec}(\sum_{l=1}^L \mathbf{C}_l^{c^l_i})$.

To improve codebook utilization and prevent collapse, we replace the argmin assignment with an optimal transport-based Sinkhorn algorithm \cite{zhang2024preventinglocalpitfallsvector}, which encourages a more uniform distribution of code usage. Furthermore, to ensure that the generated IDs retain semantic coherence, we introduce an additional contrastive objective. Using the same positive pairs $\mathcal{P}^+$, we apply an auxiliary InfoNCE loss to the reconstructed vectors $\hat{\mathbf{z}}_i$ and $\hat{\mathbf{z}}_j$. The final training objective for the RQ-VAE is:
\begin{equation}
  \mathcal{L}_{\text{RQ-VAE}} = \lambda_{0} \mathcal{L}_{\text{reconstruct}} + \lambda_{1}\mathcal{L}_{\text{commit}} +  \mathcal{L}_{\text{contrast}}(\hat{\mathbf{z}}_i, \hat{\mathbf{z}}_j) 
\end{equation}

Following this paradigm, SaviorEnc yields a behavior-aware multimodal embedding and a residual semantic ID for each item, which together serve as input features for the ranking model.

\subsection{Modal-Behavior Alignment Module}
\label{sec:3_3}
After obtaining the preliminary behavior-aligned item multimodal embedding, it is crucial to design an effective way to integrate it into the ranking model. However, since the item multimodal features extracted by the SaviorEnc are computationally expensive to update in real time, the discrepancy between the behavior space of the ranking model and the semantic space of the multimodal embeddings tends to grow as the ranking model is trained and updated. Therefore, we propose a Modal-Behavior Alignment (\textbf{MBA}) module to enhance the alignment between the semantic and behavior space.

For each item $i$, MBA module takes raw item multimodal embedding $\mathbf{z}$ and RQ code $\mathbf{c}=[c_1,\ ...,\ c_L]$ mentioned in \ref{sec:3_2} as input (subscript $i$ is omitted for simplicity). Residual quantization (RQ) is a method compressing embedding vectors into hierarchical cluster IDs, where each layer of IDs captures semantic information coarse to fine. Based on the RQ code, MBA module constructs a behavior-alignment vector, which serves as a complementary signal to bridge the gap between the frozen multimodal embedding and the dynamically updated ranking model. In practice, we firstly construct a trainable zero-initialized MBA codebook $C=[C_1,\ ...,\ C_L]$ with the same shape as the RQ codebook obtained in \ref{sec:3_2}, and each RQ code $c_l \in \mathbf{c}$ corresponds to a embedding vector $\mathbf{v}_l$ at the $l_{th}$ layer of MBA codebook. 
\begin{equation}
    \mathbf{v}_l = C_l^{c_l}
\end{equation}
\begin{equation}
    \mathbf{v} = [\mathbf{v}_1,\ ...,\ \mathbf{v}_L]
\end{equation}
After we gather the embedding vectors $\mathbf{v}=[\mathbf{v_1},\ ...,\ \mathbf{v_L}]$ from MBA codebook, we propose a way different from traditional RQ to reconstruct multimodal embedding. Following the practice\cite{rqvae,rqaudio} in image and audio domains, recent researches \cite{unified,enhancing} directly sum the embedding vectors in $\mathbf{v}$ to produce the multimodal feature. However, this approach can lead to instability in codebook training, since the gradients back-propagated to $\mathbf{v_l}$ for each $l \in \{1,...,L\}$ are identical, whereas the semantic codes follow a coarse-to-fine hierarchy and thus have different levels of importance and numeric scales. Therefore, we propose a fusion layer to adaptively learn the importance of embeddings from each residual layer, further enabling soft optimization of the codebook during gradient back-propagation. In practice, we first concatenate each embedding vector, then apply L2 normalization and use an MLP layer to fuse embeddings from each residual layer and map the result back to the original dimension. 
\begin{equation}
    \mathbf{v}_{align} = MLP(Concat([\mathbf{v}_1,\ ...,\ \mathbf{v}_L]))
\end{equation}
The output embedding vector of the MLP layer is further added to the original multimodal embedding $\mathbf{z}$, which serves as a residual signal to align multimodal embedding to behavior space. 
\begin{equation}
    \mathbf{z}_{align} = \mathbf{z} + \mathbf{v}_{align}
\end{equation}
By combining the codebook-based alignment signal with the original embedding in a skip-connect pattern, it is possible to avoid the loss of multimodal information extracted in \ref{sec:3_2}, while also helping the codebook focus on the most important content. The final output $\mathbf{z_{align}}$ of MBA module will serve as behavior-aligned multimodal embedding of candidate item or items in user behavior sequence and be fed into the bi-directional target attention module.

With the help of the trainable MBA codebook, the multimodal embedding can be updated along with the training of the whole ranking model, overcoming the restrictions caused by the semantic-behavior gap.

\subsection{Bi-Directional Target Attention Mechanism}
For the behavior feature embeddings including item IDs and statistical features $\mathbf{h}_{seq}=[\mathbf{h}_1,\ ...,\ \mathbf{h}_n]=Concat([\mathbf{ID}_1,\ ...,\ \mathbf{ID}_n],[\mathbf{S}_1,\ ...,\ \mathbf{S}_n])$ and multimodal embeddings $\mathbf{z}_{seq}=[\mathbf{z}^{align}_1,\ ...,\ \mathbf{z}^{align}_1]$ of user interaction sequence, we propose a bi-directional target attention mechanism to fuse information from both the modal-side and the behavior-side, further to extract user interests. This promotes interaction between the behavior embedding and multimodal embedding, thus enhancing the representation power of the model.

Target attention(TA) mechanism\cite{din} calculates a similarity score between user sequence and candidate item. Then the embeddings in the behavior sequence are aggregated based on similarity scores to extract the user's interest.
\begin{equation}
     TA(Q,K,V)=(\frac{(\mathbf{Q}\mathbf{W}^q)(\mathbf{K}\mathbf{W}^k)^T}{\sqrt{d}})(\mathbf{V}\mathbf{W}^v)
\end{equation}
Our bi-directional target attention mechanism consists of four TA blocks, including two regular TA blocks which are applied separately to the behavior embedding and the multimodal embedding, in order to extract user interest in the behavior space and the semantic space. This process can be formulated as follows, where subscript $cand$ represents the candidate item, and $m$ and $b$ refer to modal and behavior respectively. 

\begin{align}
    &\mathbf{h}_{b} = TA(\mathbf{h}_{cand},\mathbf{h}_{seq},\mathbf{h}_{seq})\\
    &\mathbf{h}_{m} = TA(\mathbf{z}_{cand},\mathbf{z}_{seq},\mathbf{z}_{seq})
\end{align}

The other two TA blocks aim to fuse the modal and behavior embeddings and capture cross-information. At Modal2Behavior TA block, we use the similarity score in semantic space to aggregate behavior embedding sequence. Similarly, we use the similarity score in behavior space to aggregate multimodal embedding sequence at Behavior2Modal TA block.
\begin{align}
    &\mathbf{h}_{m2b} = TA(\mathbf{z}_{cand},\mathbf{z}_{seq},\mathbf{h}_{seq})\\
    &\mathbf{h}_{b2m} = TA(\mathbf{h}_{cand},\mathbf{h}_{seq},\mathbf{z}_{seq})
\end{align}

Finally, we concatenate the outputs of the TA blocks to form the feature vector for the CTR prediction network.
\begin{align}
    &\mathbf{h}_{BiD-TA} = Concat([\mathbf{h}_{b},\mathbf{h}_{m},\mathbf{h}_{b2m},\mathbf{h}_{m2b}]) \\
    &pCTR = DNN(\mathbf{h}_{BiD-TA},other\_features)
\end{align}
We optimize the CTR prediction model, including the MBA module and bi-directional attention mechanism, by cross-entropy loss:
\begin{equation}
    Loss = -\frac{1}{N}\sum_{i=1}^N[y_i\log(pCTR_i)+(1-y_i)\log(1-pCTR_i)]
\end{equation}
where $N$ is batch size, $y_i\in\{0,1\}$ is the label of sample $i$, and $pCTR$ is the output of CTR prediction model.

\section{Experiment}
To assess the effectiveness of SaviorRec, we conduct experiments around the following key research questions.
\begin{itemize}
    \item \textbf{RQ1}: How does our model perform compared to other methods that leverage multimodal information?
    \item \textbf{RQ2}: What is the impact of our model's different components on overall performance, and how does it address the limitations of existing methods?
    \item \textbf{RQ3}: How can we achieve a trade-off between the parameter size and effectiveness of multimodal modeling?
    \item \textbf{RQ4}: What roles do behavioral and semantic information play in user interest modeling and item recommendation?
\end{itemize}

\begin{table}
  \caption{The statistic of evaluation dataset (\%). PV refers to Pages View of a candidate item.}
  \label{tab:dataset_stat}
  \begin{tabular}{cccc}
    \toprule
    \textbf{PV Group}&\textbf{Samples}&\textbf{Clicks}&\textbf{Items}\\
    \midrule
    $[0,100)$&2.24&2.16&31.07\\
    $[100,500)$&17.09&17.74&33.98\\
    $[500,1000)$&32.29&31.01&24.27\\
    $[1000,5000)$&24.90&22.39&8.74\\
    $[5000,10000)$&8.66&8.79&0.87\\
    $[10000,20000)$&14.15&16.75&0.68\\
    $[20000,\infty)$&0.67&1.16&0.39\\
  \bottomrule
\end{tabular}
\end{table}

\begin{table*}[t]
  \caption{The AUC(\%, $\uparrow$) results of SaviorRec and baselines. The best is denoted in bold font.}
  \label{tab:main_result}
  \centering
  \begin{tabularx}{\textwidth}{cc*{8}{Y}}
    \toprule
    \multirow{2}{*}{\textbf{Methods}} & \multirow{2}{*}{\textbf{Total AUC}} & \multicolumn{7}{c}{\centering \textbf{AUC across item PV Buckets}} \\
    \cmidrule(lr){3-9}
    & & \textbf{[0,100)} & \textbf{[100,500)} & \textbf{[500,1000)} & \textbf{[1000,5000)} & \textbf{[5000,10000)} & \textbf{[10000,20000)} & \textbf{[20000,$\infty$)} \\
    \midrule
    Base&71.28&70.34&70.16&70.67&71.12&73.47&72.01&71.93\\
    \midrule
    BBQRec&71.61&71.08&70.65&71.05&71.41&73.62&72.16&71.93\\
    CHIME&71.21&70.27&70.07&70.60&71.06&73.41&71.97&71.87\\
    MIM&72.02&71.71&71.20&71.50&71.82&73.92&72.48&72.02\\
    SimTier&71.36&70.28&70.23&70.76&71.22&73.52&72.03&71.79\\
    \midrule
    SaviorRec&\textbf{72.11}&\textbf{71.87}&\textbf{71.32}&\textbf{71.61}&\textbf{71.89}&\textbf{73.95}&\textbf{72.50}&\textbf{72.04}\\
    \bottomrule
  \end{tabularx}
\end{table*}

\subsection{Experimental Settings}

\subsubsection{Metrics}
(1)\textbf{CTR Prediction}:
For CTR prediction task, we use AUC as the main evaluation metric. AUC is the probability that a positive user-item pair receives a higher score than a negative user-item pair, which reflects the model's scoring ability. For a more fine-grained evaluation across items with varying popularity, we partition items according to their total historical page views (PV) and compute the AUC within each group.
(2)\textbf{Multimodal Representation Learning}: 
In addition, we directly employ a behavioral retrieval task that simulates an item-to-item (i2i) recommendation scenario to evaluate our multimodal encoder. For each historical item in a user's sequence, we retrieve the top-$K$ most similar items from the corpus using its feature vector. The ground truth consists of the items the user subsequently interacted with. We use the following standard metrics for this evaluation: Hitrate@K, which measures the proportion of ground-truth items that appear in the top-$K$ retrieved list.

\begin{table}[t]
  \caption{Ablation study of different SaviorRec components, where "w/o" is short for "without". The best is denoted in bold font.}
  \label{tab:ablation}
  \centering
  \begin{tabular}{ccc}
    \toprule
    \textbf{Methods} &\textbf{Total AUC} &$\mathbf{\Delta}$\\
    \midrule
    Base&71.28&-0.83\\
    \midrule
    w/o MBA  &72.00&-0.11\\ 
    w/o multimodal embedding &71.80&-0.31\\ 
    w/o Bi-Dirc Attn &71.98&-0.13\\
    \midrule
    SaviorRec&\textbf{72.11}&-\\
    \bottomrule
  \end{tabular}
\end{table}

\subsubsection{Baselines}
We select Taobao's cold-start model currently used online, and several studies on the application of multimodal features in recommendation systems as baselines. Considering there are significant differences among some baseline methods in the overall paradigm, we only use the corresponding modules to construct multimodal features.
\begin{itemize}
\item \textbf{Base} model is Taobao's current cold-start CTR prediction model not utilizing multimodal features; it only consists of target attention over the Item ID and other statistical features.
\item \textbf{BBQRec}\cite{bbqrec} proposes an auxiliary module within the self-attention through a non-invasive manner.
\item \textbf{CHIME}\cite{chime} designs an interest compression module that end-to end learns the user interest distribution and compresses it as a compact histogram.
\item \textbf{MIM}\cite{mim} proposes a fusion interest module to combine item ID and content interests.
\item \textbf{SimTier}\cite{deepset} computes the similarities between the candidate item and the user's interacted items, and summarizes them in a histogram to serve as a multimodal feature.

\end{itemize}

\subsubsection{Datasets}
Experiments are conducted on traffic logs collected from Taobao's homepage feed. We construct an industrial-scale cold-start dataset for training and evaluation according to Taobao's cold-start item selection algorithm, which is detailed in Tab.\ref{tab:dataset_stat}. We choose data from a three-week period in July 2025 for training, with the final day for evaluation, which contains on the order of $10^8$ samples each day.

\subsubsection{Implement Details}For the \textbf{multimodal representation learning}, we use the CN-CLIP \cite{yang2023chineseclipcontrastivevisionlanguage} architecture as our base model. We then employ a 3-layer multimodal transformer for fusing the single-model outputs. This feature generation model is trained for 10 epochs with a learning rate of 1e-4, a 1-epoch warmup period, and a batch size of 4096.

For the \textbf{semantic ID training stage}, we train the RQ-VAE module. The training runs for 150 epochs with a learning rate of 0.002 and a large batch size of 16384 to ensure stable codebook learning. The loss weights for the reconstruction objectives and commitment loss are set to $\lambda_{0}=1000$ and $\lambda_{1}=0.5$, respectively. We set the RQ codebook size of each layer $K=2048$, codebook length $L=8$, and embedding dimension $d=64$.

For \textbf{the CTR prediction model}, we set the batch size to 1024 and use AdagradV2 as the optimizer, with a learning rate that gradually decays from 0.01 to 0.001. The MBA codebook and the RQ codebook are of the same size. In the bi-directional target attention module, we use a multi-head attention mechanism \cite{multihead} with 8 heads. The final DNN part contains 3 MLP layers to predict CTR and uses LeakyReLU \cite{lrelu} as the activation function.

\subsection{Overall Performance (RQ1)}
\label{sec:overall_performance}
In this section, we compared our method with baseline recommendation models using multimodal features. The overall performance is shown in Tab.\ref{tab:main_result} and our method outperforms other baselines at both total AUC and AUC across item PV Buckets.

Compared to the base model without modality information, most methods achieve significant AUC improvement, demonstrating the substantial advantages of semantic information for user interest modeling and item recommendation. However CHIME fails to outperform the base model because its way of constructing multimodal feature embedding does not align well with the ranking model, which prevents it from capturing useful information.

Among the multimodal recommendation methods, SaviorRec performs best in all metrics, which highlights the effectiveness of the MBA module and bi-directional target attention mechanism. Compared to methods extracting statistical features such as histograms (SimTier, CHIME), methods that directly use attention mechanisms (SaviorRec, MIM) achieve better performance. This shows that attention can preserve more multimodal information and offers clear advantages. 

For AUC across different item historical page views(PVs) groups, the lower the PV, the better our method performs. This indicates that multimodal features play a more important role in the recommendation of cold-start items, due to the lack of behavioral information and as a result, insufficient modeling of item IDs.

\begin{table}[t]
\centering
\caption{Ablation study on multimodal representation learning.}
\label{tab:encoder_ablation}
\begin{tabular}{@{}l c c c@{}}
\toprule
\textbf{Model} & \textbf{Domain Adapt.} & \textbf{i2i Alignment} & \textbf{Hit@30 (\%)} \\ \midrule
\#1 (Official) &                           &                              & 28.56               \\
\#2            & \checkmark                &                              & 32.36               \\
\#3            &                           & \checkmark                   & 39.28               \\
\#4 (Ours)   & \checkmark                & \checkmark                   & \textbf{41.30}      \\ \bottomrule
\end{tabular}
\end{table}

\subsection{Ablation and Analysis (RQ2)}
We conduct ablation studies to validate the effectiveness of each component in SaviorRec as follows:
\begin{itemize}
    \item \textbf{Base}: Only consists of target attention over the Item ID and subsequent MLP layers.
    \item \textbf{w/o MBA}: We remove the MBA module and directly feed the frozen multimodal embeddings into bi-directional attention block.
    \item \textbf{w/o multimodal embedding}: We initialize the codebook in MBA module with that obtained in the RQ stage, and do not sum the codebook embedding with the original multimodal embedding.
    \item \textbf{w/o Bi-Dirc Attn}: We remove the bi-directional attention block and only apply target attention over multimodal features.
\end{itemize}

The experiment results are shown in Tab.\ref{tab:ablation}. (1) Our MBA module can add a dynamic alignment signal to the static multimodal embedding, resulting in a 0.11\% improvement in AUC. However, when we initialize the codebook in the MBA module with that obtained in the RQ stage, instead of zero-initializing it and summing the codebook embedding with the original multimodal embedding, the AUC decreases by 0.31\%. This demonstrates that the skip-connect structure we designed in the MBA module helps maintain a balance between the multimodal features and the modal-behavior alignment signal, preventing the behavioral signal from overwhelming the original multimodal information during training. (2) As we remove the bi-directional attention block, total AUC decreases by 0.13\%,  indicating that the mutual fusion between semantic and behavioral signal is beneficial for user-item modeling. 

\begin{figure}[htbp]
  \centering
  \includegraphics[width=0.85\linewidth]{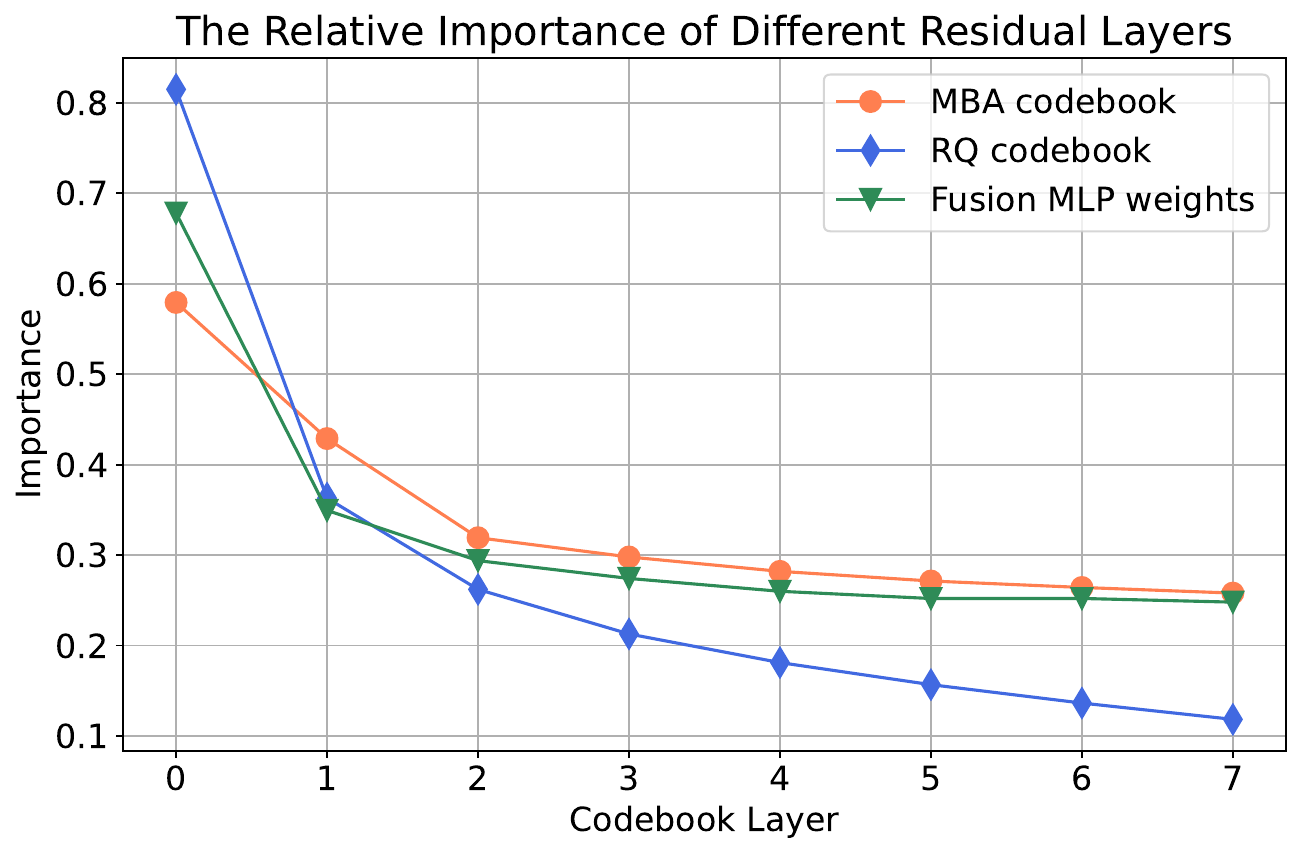}
  \caption{The relative importance of different residual layers. The importance scores are obtained by normalizing the average L2 norm of the corresponding parameters.}
  \label{fig:4_3}
\end{figure}

To verify that our zero-initialized MBA codebook can effectively learn alignment signal, we analyze the average L2 norm of codebook embeddings and subsequent fusion MLP weights of each residual layer in Fig.\ref{fig:4_3}. The MBA codebook is capable of learning the relative importance across different layers, exhibiting a coarse-to-fine feature hierarchy. In the fusion MLP layer, the network weights corresponding to each residual layer also show a similar importance distribution. This ensures that, during training, the gradients of each codebook layer are scaled in accordance with their importance, thereby avoiding the training instability caused by direct sum pooling of the codebook.

Additionally, we also conduct an ablation study on SaviorEnc's key components. We evaluate the quality of the representations using a behavioral retrieval task that simulates a real-world item-to-item (i2i) recommendation scenario. For each user, we take their historical clicked items as queries. We then leverage the corresponding multimodal representations to retrieve the top-30 most similar items from the entire item corpus. A retrieval is considered successful (as a "hit") if the retrieved list contains the item the user clicked after the query item. The results of our ablation study are presented in Table \ref{tab:encoder_ablation}. The results demonstrate the effectiveness of our approach. Domain adaptation (\#2 vs. \#1) yields a notable improvement, while i2i behavior alignment (\#3 vs. \#1) proves more critical, delivering a 10.72\% absolute gain in Hit@30. Our full SaviorEnc (\#4) achieves the best performance by combining these two complementary strategies, validating the overall design.
\subsection{Parameter Analysis of MBA Module Codebook (RQ3)}

In this section, we reduce the embedding dimension of the MBA codebook to investigate whether a balance can be achieved between model performance and the number of codebook parameters. The result is shown in Tab.\ref{tab:4_4_result}. In our original MBA Module, the embedding dimension is set to 64, consistent with the codebook used in the RQ stage. We first reduce the embedding dimension to 32 and 16, and the AUC only decreases slightly. However, when the dimension is further reduced to 8, the AUC drops more noticeably, approaching the performance of the model which only uses frozen multimodal embedding without the MBA module, but still better than baselines in \ref{sec:overall_performance}. 
These results indicate that our model can be made more lightweight with minimal performance loss, which is valuable for deployment in industrial settings. However, reducing the codebook dimension too much will constrain the modeling capability.
\begin{table}
  \caption{AUC(\%) with different codebook dimensions}
  \label{tab:4_4_result}
  \begin{tabular}{ccccc}
    \toprule
    \textbf{MBA Codebook Dimension} & 64 & 32 & 16 & 8 \\
    \midrule
    \textbf{Total AUC}     & \textbf{72.11}   &  72.07  & 72.08   & 72.03  \\
    \bottomrule
  \end{tabular}
\end{table}

\begin{figure}[htbp]
  \centering
  \includegraphics[width=0.88\linewidth]{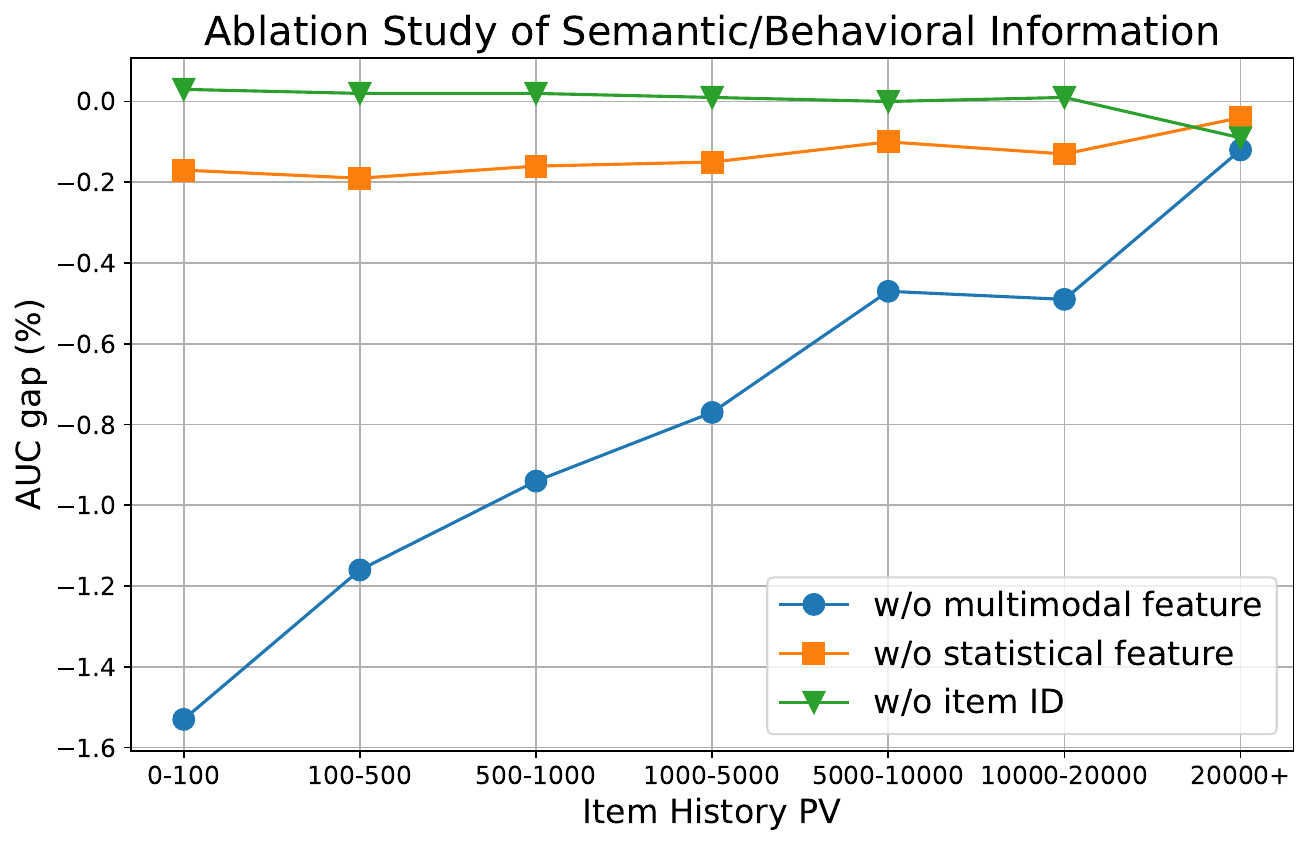}
  \caption{The AUC drop after removing ID, statistical and multimodal features.}
  \label{fig:4_5}
\end{figure}

\subsection{Effectiveness of Behavioral and Semantic Information (RQ4)}
In SaviorRec, we use ID embedding, statistical features and multimodal features to model both candidate item and items in user behavior sequence. As shown in Fig.\ref{fig:4_5}, we separately remove these features to explore the role of behavioral and semantic information in recommendation.

When we remove all multimodal features and the corresponding modules in our method, the AUC drops significantly across all PV intervals, demonstrating the importance of semantic information for cold-start scenario. For items with lower PV, the lack of user interactions leads to sparse statistical features and insufficient training of item IDs, therefore removing multimodal features has a greater negative impact.

For features in the behavior space, removing the statistical features from the bi-directional attention module leads to a smaller drop in AUC, and are equally important for both popular and cold-start items. However, item ID contributes little to cold-start recommendation and even has a negative impact when PV $<$ 5000, indicating that when the item PV is low, training the item ID embedding with a small number of samples will only lead to unstable representations. When item PV $>$ 20000, the drop in AUC without item ID suggests that it is able to effectively capture behavioral information for these popular items.

\begin{figure}[htbp]
  \centering
  \includegraphics[width=\linewidth]{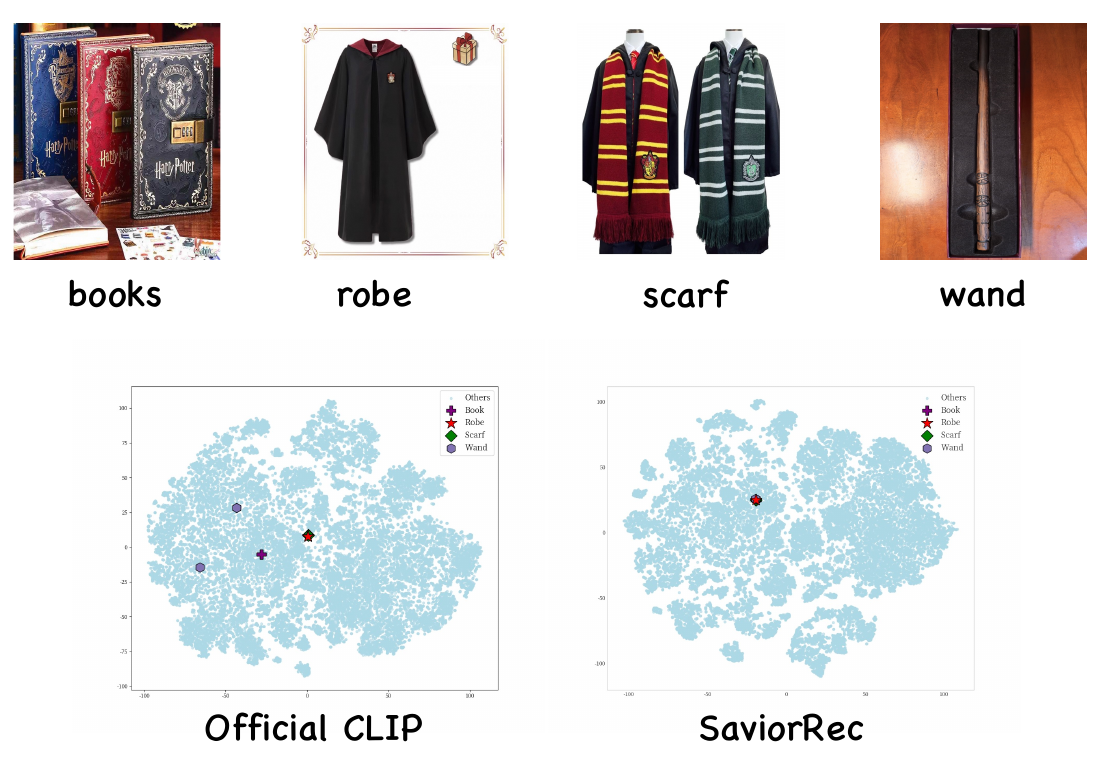}
  \caption{Visualization of item multimodal embedding space. SaviorRec achieves alignment among items of different categories under the same behavioral paradigm. }
  \label{fig:visualization}
\end{figure}

Fig.\ref{fig:visualization} visualizes the multimodal embedding spaces using t-SNE, with several \textit{Harry Potter} themed items (books, robes, scarves, wands) highlighted. The baseline ``Official CLIP'' model fails to group items with similar use interaction pattern. Our ``SaviorRec'', trained with user behavior signals, forms a single cluster for all \textit{Harry Potter} items. This demonstrates that the behavior alignment mechanism in SaviorEnc and MBA module can capture user interests.

\begin{table}
  \caption{Results of online A/B test.}
  \label{tab:4_5_abtest}
  \begin{tabular}{cccc}
    \toprule
\textbf{Metrics} & Clicks & Orders & CTR\\
    \midrule
    \textbf{Impr.(\%)} & 13.31 & 13.44 & 12.80 \\
    \bottomrule
  \end{tabular}
\end{table}

\subsection{Online A/B Test}
We deployed SaviorRec and conducted A/B test on Taobao's “Guess You Like” service, specifically targeting cold-start items identified by predefined rules. Clicks, orders, and CTR are adopted as the main evaluation metrics, since they are the primary indicators of interest in our industrial setting. Tab.\ref{tab:4_5_abtest} shows the online results, where our method achieved a 13.31\% increase in clicks, a 13.44\% increase in orders, and a 12.80\% improvement in CTR. This highlights the importance of our method in modeling cold-start items and enabling accurate recommendations.

\section{Conclusion}
In this paper, we propose SaviorRec, a novel and deployable multimodal recommendation framework that achieves alignment between semantic information and user behavior. SaviorEnc leverages co-click item pairs to generate behavior-aware multimodal representation and semantic ID. The MBA module achieves continuous alignment between semantic information and user behavior throughout the ranking model's training by dynamically updating a residual codebook. The bi-directional target attention mechanism promotes the fusion of behavior and multimodal features, and enhances the model's representation power. We conducted extensive experiments to demonstrate that SaviorRec outperforms existing methods for multimodal recommendation, verifying the effectiveness of our method. The substantial gains in online clicks, orders, and CTR demonstrate that our model is capable of utilizing behavior-aligned multimodal information for cold-start item modeling and high-quality recommendation in real-world industrial settings.


\bibliographystyle{ACM-Reference-Format}
\bibliography{sample-base}

\end{document}